# Scaling property of variational perturbation expansion for general anharmonic oscillator with $x^p$-potential

W. Janke[1,2] and H. Kleinert[2]

[1] *Institut für Physik, Johannes Gutenberg-Universität Mainz, Staudinger Weg 7, 55099 Mainz*

[2] *Institut für Theoretische Physik, Freie Universität Berlin, Arnimallee 14, 14195 Berlin*

(February 17, 1995)

## Abstract

We prove a powerful scaling property for the extremality condition in the recently developed variational perturbation theory which converts divergent perturbation expansions into exponentially fast convergent ones. The proof is given for the energy eigenvalues of an anharmonic oscillator with an arbitrary $x^p$-potential. The scaling property greatly increases the accuracy of the results.

Typeset using REVTEX

In a series of recent papers and a textbook it has been demonstrated that divergent perturbation expansions in quantum mechanics can be turned into exponentially fast convergent ones with the help of *variational perturbation theory* [1–3]. The purpose of this note is to show that the calculation can be greatly simplified by observing that the extrema of the energy depend universally, i.e., independently of the coupling strength, upon a simple scaling variable $\sigma$. Since $\sigma$ is a polynomial of degree $p/2 + 1$ in the variational parameter, this reduces the degree of the polynomials to be extremized by this factor (3 for the quartic potential). The the order of the approximation can therefore be raised by a factor $p/2 + 1$. With the exponentially fast convergence, this leads to a great increase in the accuracy of the results.

Consider a general symmetric anharmonic oscillator with a potential

$$V(x) = \frac{\omega^2}{2}x^2 + gx^p \qquad (\omega^2, g > 0), \qquad p = \text{even}. \tag{1}$$

The standard Rayleigh-Schrödinger perturbation theory yields for each energy level a power-series expansion in the dimensionless reduced coupling constant $\hat{g} = g/\omega^{\frac{p+2}{2}}$:

$$E(g) = \omega \sum_{l=0}^{\infty} e_l^{\text{BW}} \left( \frac{g}{\omega^{\frac{p+2}{2}}} \right)^l. \tag{2}$$

The expansion coefficients $e_l^{\text{BW}}$ are rational numbers which can easily be computed by a recursion scheme derived a long time ago by Bender and Wu [4]. A direct summation of the series (2) is meaningless due to the factorial growth of the coefficients $e_l^{\text{BW}}$ giving rise to a vanishing radius of convergence. An approximate evaluation is only possible at very small $g$ after truncating the series at an order $l \approx 1/g$.

An exponentially fast convergent evaluation of the series (2) becomes possible for all coupling strengths [3] with the help of variational perturbation theory. It is here where the restriction to even powers $p$ in the potential (2) is necessary; otherwise a second variational parameter is required.

The variational perturbation expansion is derived as follows. First, the harmonic term of the potential is split into an arbitrary harmonic term and a remainder

$$\frac{\omega^2}{2}x^2 = \frac{\Omega^2}{2}x^2 + \left(\frac{\omega^2}{2} - \frac{\Omega^2}{2}\right)x^2. \tag{3}$$

Rewriting

$$V(x) = \frac{\Omega^2}{2}x^2 + V_{\text{int}}(x), \tag{4}$$

with an interaction

$$V_{\text{int}}(x) = g(rx^2 + x^p), \quad r = (\omega^2 - \Omega^2)/2g, \tag{5}$$

one performs a perturbation expansion in powers of $g$ at a fixed $r$,

$$E_N(g, r) = \Omega \sum_{l=0}^{N} e_l(r) \left(\frac{g}{\Omega^{\frac{p+2}{2}}}\right)^l. \tag{6}$$

The calculation of the reexpansion coefficients $e_l(r)$ up to a specific order $N$ does not require much additional work since it is easily obtained from the ordinary perturbation series (2) by replacing $\omega$ by $\sqrt{\Omega^2 + 2gr}$ and expanding in powers of $g$ up to the $N$th order. This yields

$$e_l = \sum_{j=0}^{l} e_j^{\text{BW}} \binom{(1 - \frac{p+2}{2}j)/2}{l-j} (2r\Omega^{\frac{p-2}{2}})^{l-j}. \tag{7}$$

The truncated power series

$$W_N(g, \Omega) := E_N\left(g, (\omega^2 - \Omega^2)/2g\right) \tag{8}$$

is certainly independent of $\Omega$ in the limit $k \longrightarrow \infty$. At any finite order, however, it does depend on $\Omega$, the approximation having its fastest speed of convergence where it depends least on $\Omega$, e.g., at points where $dW_N/d\Omega = 0$. If we denote the order-dependent optimal value of $\Omega$ by $\Omega_N$, the quantity $W_N(g, \Omega_N)$ is the new approximation to $E(g)$.

The extremization yields a large number $3N$ of $\Omega$-values from which to select the best $\Omega_N$ by choosing the flattest extremum, i.e., the one with the smallest second derivative. We shall now demonstrate that the scaling variable

$$\sigma \equiv -2r\Omega^{\frac{p-2}{2}} = \frac{\Omega^{\frac{p-2}{2}}(\Omega^2 - \omega^2)}{g} \tag{9}$$

makes the extremality condition $dW_N/d\Omega = 0$ a universal function of $\sigma$:

$$P_N(\sigma) = 0, \tag{10}$$

where $P_N$ is a polynomial of degree $N$ in $\sigma$. The polynomial is obtained by forming the derivative of $W_N(g, \Omega)$ and removing the $N$th power of the reduced coupling constant as follows:

$$\frac{dW_N}{d\Omega} = \left(\frac{g}{\Omega^{\frac{p+2}{2}}}\right)^N P_N(\sigma). \tag{11}$$

The condition (10) increases the order to which the expansion can be evaluated by a factor of $p/2+1$. Since the convergence is exponential, this greatly increases the accuracy which can be reached by a variational perturbation expansion. This nontrivial property of the variational approach was discovered empirically while treating the quartic anharmonic oscillator in Ref. [5] (for details see [3]), but not understood.

We prove Eq. (11) and derive the formula

$$P_N(\sigma) = -2\frac{de_{N+1}(\sigma)}{d\sigma}. \tag{12}$$

Together with (7), we obtain the polynomial of degree $N$:

$$P_N(\sigma) = -2\frac{de_{N+1}(\sigma)}{d\sigma} \tag{13}$$

$$= 2\sum_{j=0}^{N} e_j^{\text{BW}} \begin{pmatrix} (1 - \frac{p+2}{2}j)/2 \\ N+1-j \end{pmatrix} (N+1-j)(-\sigma)^{N-j}. \tag{14}$$

The proof proceeds as follows. Differentiating

$$W_N = \Omega \sum_{l=0}^{N} e_l(\sigma) \left(\frac{g}{\Omega^{\frac{p+2}{2}}}\right)^l \tag{15}$$

with respect to $\Omega$ we obtain

$$\frac{dW_N}{d\Omega} = \sum_{l=0}^{N} \left[e_l(\sigma) - \frac{p+2}{2}le_l(\sigma) + \Omega\frac{de_l}{d\Omega}\right]\left(\frac{g}{\Omega^{\frac{p+2}{2}}}\right)^l. \tag{16}$$

With the help of

$$\Omega \frac{de_l}{d\Omega} = \left[ 2\frac{\Omega^{\frac{p+2}{2}}}{g} + \frac{p-2}{2}\sigma \right] \frac{de_l}{d\sigma} \tag{17}$$

this becomes

$$\frac{dW_N}{d\Omega} = \sum_{l=0}^{N} \left[ \left(1 - \frac{p+2}{2}l\right) e_l(\sigma) + \left(2\frac{\Omega^{\frac{p+2}{2}}}{g} + \frac{p-2}{2}\sigma\right) \frac{de_l}{d\sigma} \right] \left(\frac{g}{\Omega^{\frac{p+2}{2}}}\right)^l. \tag{18}$$

Rearranging the sum gives

$$\begin{aligned}
\frac{dW_N}{d\Omega} &= 2\frac{de_0}{d\sigma} \left(\frac{g}{\Omega^{\frac{p+2}{2}}}\right)^{-1} \\
&+ \sum_{l=1}^{N-1} \left[ \left(1 - \frac{p+2}{2}l\right) e_l + \frac{p-2}{2}\sigma \frac{de_l}{d\sigma} + 2\frac{de_{l+1}}{d\sigma} \right] \left(\frac{g}{\Omega^{\frac{p+2}{2}}}\right)^l \\
&+ \left[ \left(1 - \frac{p+2}{2}N\right) e_N + \frac{p-2}{2}\sigma \frac{de_N}{d\sigma} \right] \left(\frac{g}{\Omega^{\frac{p+2}{2}}}\right)^N.
\end{aligned} \tag{19}$$

The first term vanishes trivially since $e_0$ does not depend on $\sigma$.

The crucial observation essential to our result is that the sum in the second line vanishes term by term:

$$\left(1 - \frac{p+2}{2}l\right) e_l + \frac{p-2}{2}\sigma \frac{de_l}{d\sigma} + 2\frac{de_{l+1}}{d\sigma} = 0. \tag{20}$$

To derive this we take the derivative of (7),

$$2\frac{de_{l+1}}{d\sigma} = 2\sum_{j=0}^{l} e_j^{\mathrm{BW}} \binom{(1 - \frac{p+2}{2}j)/2}{l+1-j} (j-l-1)(-\sigma)^{l-j}, \tag{21}$$

and use the identity

$$2\binom{(1 - \frac{p+2}{2}j)/2}{l+1-j} = \frac{\frac{p-2}{2}j + 2l - 1}{j - l - 1} \binom{(1 - \frac{p+2}{2}j)/2}{l-j} \tag{22}$$

to rewrite (21) as

$$2\frac{de_{l+1}}{d\sigma} = \sum_{j=0}^{l} e_j^{\mathrm{BW}} \binom{(1 - \frac{p+2}{2}j)/2}{l-j} \left(\frac{p-2}{2}j + 2l - 1\right)(-\sigma)^{l-j}. \tag{23}$$

Since

$$\frac{p-2}{2}\sigma\frac{de_l}{d\sigma} = \sum_{j=0}^{l} e_j^{\text{BW}} \begin{pmatrix} (1 - \frac{p+2}{2}j)/2 \\ l-j \end{pmatrix} \frac{p-2}{2}(l-j)(-\sigma)^{l-j}, \tag{24}$$

we find

$$\frac{p-2}{2}\sigma\frac{de_l}{d\sigma} + 2\frac{de_{l+1}}{d\sigma} = \left(\frac{p+2}{2}l - 1\right)e_l. \tag{25}$$

This proves, indeed, that each term in the sum of (19) vanishes identically.

Finally, inserting Eq. (20) for $l = N$ into the last term of Eq. (19), we obtain the desired result (11):

$$\frac{dW_N}{d\Omega} = -\left(\frac{g}{\Omega^{\frac{p+2}{2}}}\right)^N 2\frac{de_{N+1}(\sigma)}{d\sigma}. \tag{26}$$

The discovery of the scaling variable $\sigma$ was essential for the recent determination of the strong-coupling expansions of the energy levels of the quartic anharmonic oscillator to a great accuracy [5] (described in detail in [3]).

It will be interesting to see whether there exists similar universal scaling variables for odd interaction powers $p$ where the asymmetry of the potential requires the introduction of a second variational parameter, a shift in the average position of the harmonic trial potential in [6].

W.J. thanks the Deutsche Forschungsgemeinschaft for a Heisenberg fellowship.